\documentclass[aps,prl,twocolumn,superscriptaddress,showpacs]{revtex4}
\usepackage{graphicx}

\begin{document}

%Title of paper
\title{Quantum size effects on the perpendicular upper critical field
in ultra-thin lead films}

\author{Xin-Yu Bao}
\author{Yan-Feng Zhang}
\author{Yupeng Wang}
\author{Jin-Feng Jia}
\author{Qi-Kun Xue}
\affiliation{Beijing National Laboratory for Condensed Matter
Physics, Institute of Physics, Chinese Academy of Sciences,
Beijing 100080, China}
\author{X. C. Xie}
\affiliation{Beijing National Laboratory for Condensed Matter
Physics, Institute of Physics, Chinese Academy of Sciences,
Beijing 100080, China} \affiliation{Department of Physics,
Oklahoma State University, Stillwater, OK 74078, USA }
\author{Zhong-Xian Zhao}
\email[E-mail: ] {zhxzhao@aphy.iphy.ac.cn} \affiliation{Beijing
National Laboratory for Condensed Matter Physics, Institute of
Physics, Chinese Academy of Sciences, Beijing 100080, China}

\date{Submitted 14 July 2005, Revised 7 October 2005}

\begin{abstract}
We report the thickness-dependent (in terms of atomic layers)
oscillation behavior of the perpendicular upper critical field
$H_{c2\perp}$ in the ultra-thin lead films at the reduced
temperature ($t=T/T_c$). Distinct oscillations of the normal-state
resistivity as a function of film thickness have also been
observed. Compared with the $T_c$ oscillation, the $H_{c2\perp}$
shows a considerable large oscillation amplitude and a $\pi$ phase
shift. The oscillatory mean free path caused by quantum size
effect plays a role in $H_{c2\perp}$ oscillation.
\end{abstract}


\preprint{\Large{LG10098}}

\maketitle

% body of paper here - Use proper section commands
% References should be done using the \cite, \ref, and \label commands

%\section{Introduction}

There is a long history of scientific research on superconducting
thin films. In particular, theoretical and experimental studies
have been carried out to understand how the film thickness affects
the superconducting properties. It seems that the reported
experimental results of thin films can be explained by the
existing theories of superconductivity \cite{parks, deGennesa,
tinkhama, tinkhamb, deGennes, miller, schuller}. However, most
previously studied superconducting films were still relatively
thick, normally over several tens of nanometers, and the film
morphology was usually poor. If the film surface is atomically
uniform and the thickness is further reduced to several nanometers
so that the quantum size effects become apparent, a natural
question arises: will some unexpected new phenomena emerge? In
particular, does the conventional theory still work?

Previous theoretical works have predicted many possible prominent
physical properties modulated by quantum size effects: electronic
structure, critical temperature, electron-phonon interaction,
resistivity, Hall conductivity and so on
 \cite{blatt, mingyu, peter, sarma, trivedi, wei}. There are
also some related important experimental results \cite{komnik,
goldman, jalochowski, bozler, pfennigstorf, jalochowski2, chiang1,
chiang2}, such as the $T_c$ oscillations in ultra-thin Pb films,
which are caused by the density of states oscillations in confined
quantum well structures \cite{bao, chiang} and by the
electron-electron interaction mediated by quantized confined
phonons \cite{sarma, yfzhang}. However, the properties of the
upper critical field affected by the quantum size effect have not
been reported in previous work. In this Letter, we report our
experimental observation of the oscillatory $H_{c2\perp}$ through
magneto-transport measurement of ultra-thin Pb films. The
oscillations are similar to those of $T_c$ but the motivations are
more complex. Besides the factors for $T_c$ oscillation, we
interpret this unexpected phenomena by the oscillatory mean free
path in ultra-thin superconducting films caused by the quantum
size effect.

%\section{Experiment}

The 3mm$\times$10mm sized Si(111) wafers were used as substrates
and prepared by the standard cleaning procedure to obtain the
clean Si(111)-$7\times7$ surface. The base pressure of the
UHV-MBE-STM-ARPES (Angle Resolved Photoemission Spectroscopy)
combined system we used was about $5\times10^{-11}$ Torr. The Si
substrate was cooled down to 145K during the MBE layer-by-layer
growth of the Pb films. The growth rate was controlled at 0.2
ML/Min (Monolayer/Minute) and a RHEED (Reflection High Energy
Electron Diffraction) was used for real time monitoring of the
growth. After deposition, the sample was warmed up slowly to room
temperature and transferred to the analysis chamber where the STM
and ARPES were used to investigate the surface topography and the
electronic structures, respectively \cite{yfzhang}. For ex-situ
magneto-transport measurements, all the Pb films were covered with
a Au protection layer of 4ML before being taken out of the UHV
system.

The $R-H$ measurements were carried out in a very short time after
the samples were taken out of the vacuum. The applied field was
perpendicular to the sample surface and the temperatures were set
near and below $T_c$. To avoid trapping flux in, the magnet was
discharged to zero in oscillate mode and the sample was warmed up
to $8K$ before the $R-H$ measurement for each temperature. Then
the perpendicular upper critical field $H_{c2\perp}$ at different
temperatures was obtained from the $R-H$ measurements at the field
where the resistance reached half of the normal state resistance
$R_{N}$. The resistance approaches $R_{N}$ very gradually because
of the magnetoresistance effect. So we took $R_{N}$ as the
resistance where the resistance variation ratio is within 0.1\%.

\begin{figure}
\includegraphics[width=80mm]{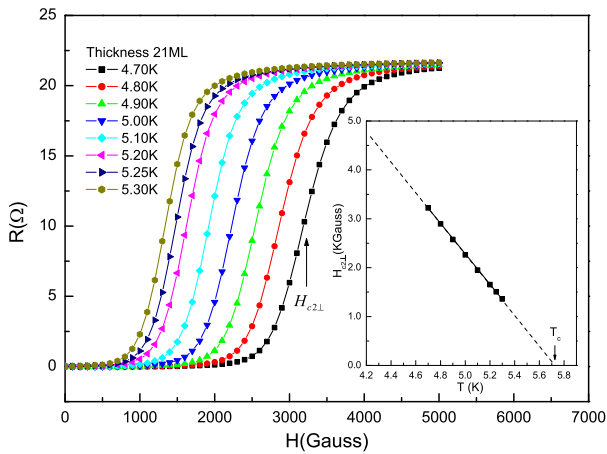}
\caption{\label{fig1}R-H curve of the 21 ML sample. The magnetic
field is perpendicular to the sample surface. The black arrow
indicates the determined upper critical field at 4.70K. The inset
shows the $H_{c2\perp}$ as a function of $T$ for this sample. The
plot is linearly extrapolated with dashed lines to both high and
low temperature sides. The measurements were carried out with a
Quantum Design Magnetic Property Measurement System(MPMS-5).}
\end{figure}

\begin{figure}
\includegraphics[width=80mm]{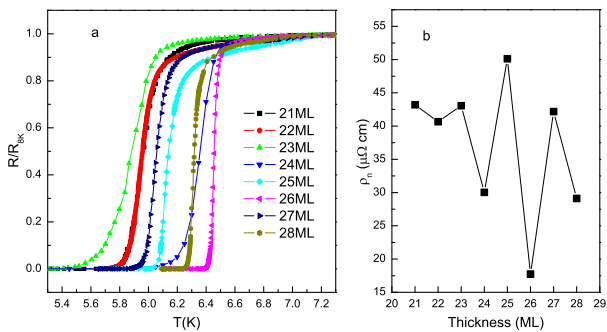}
\caption{\label{fig2} The reduced resistances of Pb films as a
function of temperature are shown in (a). The resistances are
normalized by the normal state resistance at $T=8K$. (b) shows an
oscillation of normal state resistivity at $8K$ as a function of
film thickness.}
\end{figure}
%\section{Results and Analysis}

Figure 1 shows the $R-H$ curves of a 21 ML sample at different
temperatures. The arrow points out the defined perpendicular upper
critical field $H_{c2\perp}$ at 4.7K. The inset of Fig.1 shows
$H_{c2\perp}$ vs. temperature for the 21 ML film. It shows a
perfect linear dependence on $T$ near $T_c$, which is a typical
property of a superconductor with a high value of the
Ginzburg-Landau parameter $\kappa$. The inset of Fig.1 can be used
to determine the zero field critical temperature $T_{c}$ by
extrapolating the plot to $H_{c2\perp}=0$. $T_c$ determined in
this way is shown as a function of thickness in Fig.4(a).
Normally, a direct way of determining critical temperature is
through the $R-T$ measurement at zero field. We find that the
critical temperatures determined by both methods show a consistent
oscillation behavior and the values are quite close for every
thickness.

The reduced $R-T$ curves of Pb films from 21 ML to 28 ML are shown
in Fig.2(a). The normal state resistivity $\rho_n$ oscillation
with film thickness at $T=8K$ is shown in Fig.2(b). The similar
resistivity oscillations caused by quantum size effect have been
reported in single crystalline Pb and Pb-In thin films at $T=110K$
\cite{jalochowski}. But in polycrystalline films, oscillations of
the normal state resistivity have not been observed although $T_c$
has been found to oscillate with film thickness \cite{goldman}. In
our experiment, both $T_c$ and $\rho_n$ oscillations are observed.
It indicates that the quantum size effects show up in both
superconducting state and normal state but the intensities and
mechanisms may vary in different way depending on sample
conditions.

Figure 3(a) shows $H_{c2\perp}$ as a function of the reduced
temperature $t=T/T_{c}$. For every thickness $H_{c2\perp}$ shows a
good linear dependence on $t$ near $t=1$. $H_{c2\perp}$ versus
film thickness for t=0.90 and 0.95 are shown respectively in
Fig.3(b). It is shown that with the film thickness variation,
$H_{c2\perp}$ exhibits an oscillation behavior, which is similar
to the reported $T_c$ oscillation \cite{bao}. However, the
oscillation of $H_{c2\perp}$ are $\pi$ out of phase to that of
$T_c$, i.e., peaks appear in the odd layer samples where dips
appear in the even layer samples, which is opposite to the $T_c$
oscillation shown in Fig.4(a).

In the early theories proposed to understand the magnetic
properties of thin film superconductors, the TGS (Tinkham, de
Gennes and Saint James) theory \cite{tinkhama, deGennes} was
validated as showing a good agreement with the former experimental
results \cite{miller, schuller}. According to TGS theory, the
upper critical fields $H_{c2\perp}$ near $T_{c}$ should
monotonically increase when the film thickness decreases, which
can be described in the following form \cite{miller}:
\begin{equation}
\label{1}
 H_{c\perp}(T,d)=\sqrt{2}\kappa(T,\infty)H_{c}(T)(1+b/d),
\end{equation}
where $\kappa(T,\infty)=2\sqrt{2}\pi
H_{c}(T)\lambda_{\infty}^{2}(T)/\Phi_{0}$ and
$b=3\lambda_{L}^{2}(T)\xi_{0}/8\lambda_{\infty}^{2}(T)$. Here
$H_{c}(T)$ is the thermodynamic critical field, $\lambda_{L}$ is
the London penetration depth, $\lambda_{\infty}$ is the bulk weak
field penetration depth, $\Phi_{0}$ is the flux quantum
($\Phi_{0}=hc/2e=2.07\times10^{-15}Wb$) and $d$ is the film
thickness. In Fig.3(b), the dashed lines, calculated using Eq.(1)
and the related parameters in previous work \cite{miller} with
film thicknesses appropriate to our samples, show the same
tendency as the experimental curves if the oscillations are
ignored. The measured $H_{c2\perp}$ values of our samples are
about three times larger than the calculated values (note the
different scales on the two sides of Fig.3(b)), which may be
caused by stronger interface or impurity scattering in our films
that gives rise to a large resistivity, thus large $H_{c2\perp}$
(see discussion below). The linear dependence on $t$ shown in
Fig.3(a) also gives an information that for a given film
thickness, the temperature dependence follows reasonably well with
Eq.(1) whether that particular film is at the peak or valley of
the $H_{c2\perp}$ oscillation.

\begin{figure}
\includegraphics[width=80mm]{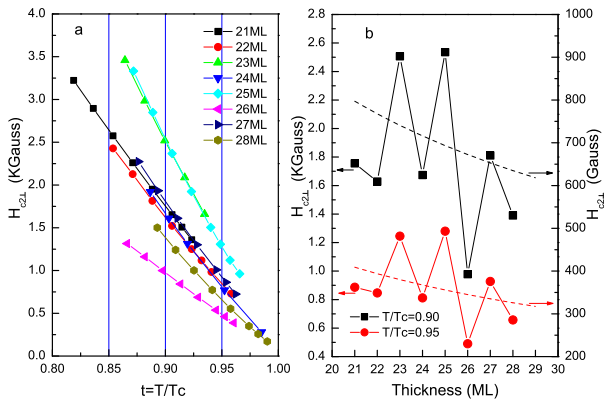}
\caption{\label{fig3} (a) shows the perpendicular upper critical
field versus the reduced temperature t. The oscillation behavior
at t=0.90 and 0.95 are plotted in (b). The dashed lines correspond
to the calculated results using Eq.(1).}
\end{figure}

\begin{figure}
\includegraphics[width=80mm]{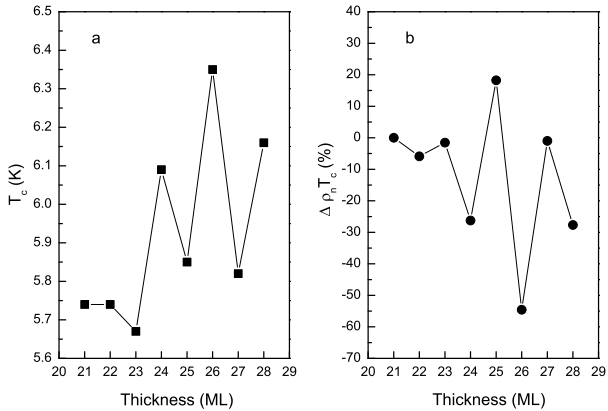}
\caption{\label{fig3} (a) shows the oscillation behavior of $T_c$
with film thickness, which is defined by the way shown in the
insert of Fig.1. The rescaled $\rho_n T_c$ variation is shown in
(b), which is defined in the following way: $\Delta \rho_n
T_c=(\rho_n T_c-\rho'_n T'_c)/\rho'_n T'_c$, where $\rho'_n T'_c$
is the value of $\rho_n T_c$ for the 21ML film.}
\end{figure}

The TGS theory above includes surface scattering effects but does
not consider the quantum size effects that occurs in ultra thin
films. The absent $H_{c2\perp}$ oscillation from TGS theory means
that the thickness depended quantum size effect is the original
source of the $H_{c2\perp}$ oscillation. According to the G-L
(Ginzburg-Landau) theory, $H_{c2\perp}$ is determined by the
in-plane coherence length $\xi_\parallel$. In a three dimensional
anisotropic superconductor, the perpendicular upper critical field
near $T_{c}$ is given by
 \cite{schuller, tinkhamb}:
$H_{c2\perp}=\frac{\Phi_{0}}{2\pi\xi_\parallel^{2}}$. Our
ultra-thin films are thinner than 10nm which is much smaller than
the Pipard coherence length of a bulk Pb superconductor
($\xi_{0}^{bulk}=83nm$), we can use the quasi two dimensional
formula \cite{sadovskii}:
\begin{equation}
\label{2}
 \left(\frac{dH_{c2\perp}}{d(T/T_c)}\right)_{T_{c}}=
 -\frac{\Phi_{0}}{2\pi\xi_\parallel^{2}}.
\end{equation}
For the linear dependence on $t$ near $t=1$ shown in Fig.3(a),
$H_{c2\perp}$ has the same oscillation behavior with thickness as
that of  $-\left(\frac{dH_{c2\perp}}{dt}\right)$ at a certain t.
The system should be considered as a dirty-limit superconductor
because of the strong scattering. For dirty superconductors near
$T_c$, $\xi_\parallel^{2} \approx \xi_0 l$ where $\xi_0$ is the
Pipard coherence length and $l$ is the mean free path for a film
\cite{deGennesa, tinkhamb}. According to BCS theory, $\xi_0
\propto 1/T_c$, therefore we can get $H_{c2\perp} \propto {T_c}/l$
at a certain $t$. In Fig.3(b) and Fig.4(a), it is shown that the
oscillation amplitude of $H_{c2\perp}$ and $T_c$ are about 40\%
and 10\% respectively. On the other hand, the mean free path $l$
and the normal state resistivity $\rho_n$ have the following
relation: $l \propto 1/{\rho_n}$, from which we can derive
$H_{c2\perp} \propto \rho_n {T_c}$. In Fig.2(b), the $\rho_n$
oscillation shows a big amplitude about 60\% and the same phase as
$H_{c2\perp}$. The rescaled variation of $\rho_n {T_c}$ is shown
as $\Delta \rho_n {T_c}$ in Fig.4(b), which fits well with the
oscillation behavior of $H_{c2\perp}$. It implies $\rho_n$
oscillation dominates over the $T_c$ oscillation in $H_{c2\perp}$
and gives rise to a $\pi$ phase shift between $T_c$ and
$H_{c2\perp}$ oscillations. In earlier works, the oscillatory
conductivity in quantized thin films has been observed
\cite{jalochowski} and the effects of impurity and surface and
interlayer roughness on quantum size effects in thin films have
been discussed \cite{trivedi, meyerovich}. Though at the moment we
do not have a complete answer to the oscillations of $\rho_n$ with
thickness for our films with atomically uniform surfaces, the
previous experiments on the layer-spacing
oscillation\cite{chiang1} provides a strong indication that the
modulation of the interface roughness with thickness may play an
important role. In that experiment they found that the interlayer
spacings oscillate with a period of quasi double layer and
even-monolayer samples have shorter interlayer spacings. This is
also supported by the binding energy modulation observed
\cite{bao, yfzhang}. It indicates the lattice feels at home with
the conduction electrons for the even-monolayer samples while it
is not so for the odd-monolayer samples. The unaccommodating
lattice and conduction elections in the odd-layer samples could
induce some lattice distortion and therefore enhance the interface
roughness. This enhanced interface roughness must induce a higher
resistivity.

%\section{Discussion and Conclusion}

We believe the experimental findings in our ultra-thin films are
 due to a variety of  combined quantum size effects from ultra-thin film
thickness. The quantum size effect can show up either as a
modulation of the interface roughness induced by the interlayer
spacings as well as the modulation of the phonon modes and the
electron-phonon couplings which both affect the normal state
transport properties of the samples, of course, also causing the
wave vector quantization along the thickness direction.  Under the
circumstance, only the components of electronic wave vector in the
surface plane, i.e., the x-y plane, have a continuous
distribution. Therefore, the electron density distribution is
rather inhomogeneous along the z direction. The modulation of the
electron-densities may further feedback to the electron-interface
and electron-phonon scattering processes and therefore to the mean
free path. Another relevant issue is that the G-L theory is only a
mean-field theory in which all the short-distance fluctuations are
integrated out. For our ultra-thin films, to give an adequate
description of all the electronic states and the scattering
processes, we must go back to the microscopic theory of BCS
superconductivity within the subband framework and derive the
multi-band G-L theory. The G-L order parameter $\Psi$
perpendicular to the film is limited to quantized values and may
also show modulation with the interlayer-spacing modulation. Each
subband may have a different value of coherence length $\xi$ in
the x-y plane, namely $\xi_{n,d}$, where $n$ is the subband index
and $d$ is the number of the monolayers. In general, $H_{c2\perp}$
is determined by a matrix equation with $m$ being the size of the
matrix in which $m$ is the number of subbands below the Fermi
energy. In the limit that one of the $\xi_{n,d}$ is much smaller
than all the others, $H_{c2\perp}$ is predominantly determined by
this minimum value, which could be much higher than that of the
bulk. The story here is similar to that of the newly discovered
superconductor $MgB_{2}$, where only two bands are involved
\cite{MgB2}. If the film becomes thicker, the number of subbands
will increase. The interaction of subbands will weaken the quantum
size effects and the coherence length will be close to the average
one. The oscillation behavior of $H_{c2\perp}$ will eventually
disappear beyond a large thickness.

In conclusion, a large oscillation  of $H_{c2\perp}$ in the ultra
thin lead films are observed as a function of film thickness. The
$H_{c2\perp}$ oscillation is opposite to that of $T_c$ in phase
and cannot be simply attributed to the modulation of the density
of states and $T_c$. A large value of $H_{c2\perp}$ is also
observed. Considering the interface and surface scattering and the
modulation of coherence length and mean free path induced by the
quantum size effect, a possible mechanism is proposed to explain
both the anomalous oscillation of $H_{c2\perp}$ and its large
value. We believe that a quantitative description for the findings
in our experiments must be based on the combined quantum size
induced modulation effects on the interlayer structures,
electronic structures, phonons, electron-phonon and
electron-interface scattering processes. Further consideration
about the flux dynamics is also necessary  by including the
interface and surface scattering effects and two-dimensional
fluctuations in the multi-band G-L theory.

%\section{Acknowledgments}

\begin{acknowledgments}
We are grateful to Professors Lu Yu,  Hai-Hu Wen, Dong-Ning Zheng,
Jue-Lian Shen, Michal Ma and Li Lu for useful discussions. We
thank Shun-Lian Jia, Wei-Wen Huang and Hong Gao for their help in
measurement. We also thank Tie-Zhu Han and Zhe Tang for their help
in sample preparation. This work was supported by the National
Science Foundation, the Ministry of Science and Technology of
China and the Knowledge Innovation Project of the Chinese Academy
of Sciences. X.C. Xie is supported by US-DOE under Grant No.
DE-FG02-04ER46124 and NSF-MRSEC under DMR-0080054.
\end{acknowledgments}

% Create the reference section using BibTeX:

\end{document}